\documentclass[a4paper]{article}
\usepackage[latin1]{inputenc}
\usepackage[T1]{fontenc}
\usepackage[english]{babel}
\usepackage{amsmath,amssymb,amsfonts}
\usepackage[margin=2cm]{geometry}
\usepackage{array}
\usepackage[pdftex]{graphicx}

\title{Toward tunable RNA thermo-switches for temperature dependent gene expression}
\author{ Oscar M.J.A. Stassen$^{1}$,  
Ruud J.J.Jorna$^{1}$,  
Bastiaan A. van den Berg$^{2}$, 
Rad Haghi$^{3}$, \and
Farzad Ehtemam$^{3}$, 
Steven M. Flipse$^{1}$, 
Marco J.L. de Groot$^{1,2}$, 
Janine A. Kiers$^{1,4}$ \and  
I. Emrah Nikerel $^{1}$
Domenico Bellomo$^{1,2}$ \thanks{Corresponding author. E-mail: domenico.bellomo@gmail.com}\and
{\normalsize
\begin{tabular}{p{17cm}}
$^{1}$ Department of Biotechnology, Faculty of Applied Sciences, Kluyver Centre for Genomics of Industrial Fermentation, Delft University of Technology, Julianalaan 67, 2628 BC Delft, The Netherlands \\
$^{2}$ Delft Bioinformatics Lab, Faculty of Electrical Engineering, Mathematics and Computer Science, Delft University of Technology, Mekelweg 4, 2628 CD, Delft, the Netherlands\\
$^{3}$ Faculty of Mechanical, Maritime and Materials Engineering, Delft University of Technology, Mekelweg 2, 2628 CD Delft, The Netherlands\\
$^{4}$ Department of Philosophy, Faculty of Technology, Policy and Management Delft University of Technology, Jaffalaan 5, 2628 BX Delft Delft, The Netherlands
\end{tabular}}}

\date{}
\begin{document}

\maketitle

\section*{Abstract}
\paragraph{Background}
RNA thermometers are mRNA strands with a temperature dependent secondary structure: depending on the spatial conformation, the mRNA strand can get translated (on-state) or can be inaccessible for ribosomes binding (off-state).These have been found in a number of micro{}-organisms (mainly pathogens), where they are used to adaptively regulate the gene expression, in response to changes in the environmental temperature. Besides naturally occurring RNA thermometers, synthetic RNA thermometers have been recently designed by modifying their natural counterparts \cite{Hofacker2003}. The newly designed RNA thermometers are simpler, and exhibit a sharper switching between off- and on-states. However, the proposed trial-and-error design procedure has not been algorithmically formalized, and the switching temperature is rigidly determined by the natural RNA thermometer used as template for the design.

\paragraph{Results}
We developed a general algorithmic procedure (consensus distribution) for the design of RNA thermo-switches with a tunable switching temperature that can be decided in advance by the designer. A software tool with a user friendly GUI has been written to automate the design of RNA thermo{}-switches with a desired threshold temperature. Starting from a natural template, a new RNA thermometer has been designed by our method for a new desired threshold temperature of 32C. The designed RNA thermo{}-switch has been experimentally validated by using it to control the expression of lucifarase. A 9.2 fold increase of luminescence has been observed between 30C and 37C, whereas between 20C and 30C the luminescence increase is less than
3-fold. 
\paragraph{Conclusions} 
This work represents a first step towards the design of flexible and tunable RNA thermometers that can be used for a precise control of gene expression without the need of external chemicals and possibly for temperature measurements at a nano{}-scale resolution.

\section*{Background }
\subsection*{Synthetic biology}

Over the past few years, synthetic biology has emerged as a new research field. Its rapid growth is mainly due to advances in the experimental techniques for DNA sequencing and synthesis, and to the application of engineering approaches and methods to biology. The key idea of synthetic biology is that complex biochemical systems can be designed and operated in living cells by combining simple \textit{building blocks} with well defined functionalities and standard input-output behaviour \cite{Endy2005}. The Registry of Standard Biological Parts \cite{Registry} maintained by the Massachusetts Institute of Technology and yearly fuelled by iGEM competition \cite{igem2008}, currently represents the main international effort to build an open source repository of well characterized genetic building blocks.   

In the last few years, the toolbox available to the synthetic biologist has been greatly enhanced by the inclusion of several RNA{}-based building blocks (e.g. ribo-switches, ribo-regulators, RNAi) acting at the level of post transcriptional regulation \cite{Serganov2007, Carthew2009, Kim2007, Narberhaus2006}. RNA thermometers are a quite recent entry in this class of genetic devices. The growing interest in the analysis and design of RNA thermometers is triggered by the perspective of efficiently controlling gene expression just by means of the growth temperature without adding expensive or harmful chemicals \cite{Waldminghaus2008}, and by the need to accurately measure temperature with high-spatial resolution, as demanded by new technologies like integrated circuits lithography or thermal imaging \cite{Lee2007}. 

\subsection*{RNA thermometers}
\begin{figure}[h]
\centering
\includegraphics[width=0.75\textwidth]{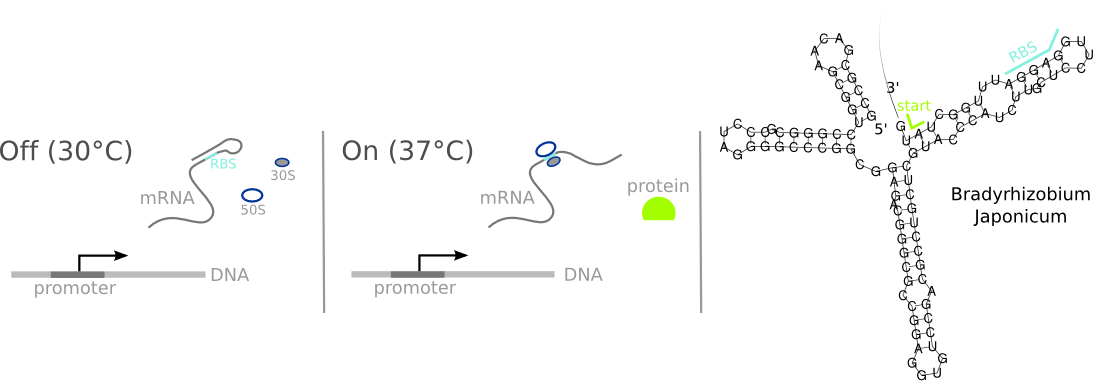}
\caption{Illustration of the working principle of an RNA thermometer. (left and middle panel) The RNA thermometer in figure allows translation at 37C, but not at 30 C. Indicated with 30S and 50S are the subunits of the ribosome. (right panel) Example of a natural occurring RNA thermometer, found in the bacterium Bradyrhizobium japonicum. RBS: Ribosome Binding Site; Start: translation start site.}
\label{fig:Figure1}
\end{figure}

RNA thermometers are mRNA strands with a temperature{}-dependent secondary structure: if the temperature is below a certain threshold, the RNA folds forming an hairpin loop , in such a way that the ribosome binding site (RBS) gets occluded, thus preventing the translation. As soon as the temperature rises above the threshold, the base pairings get destabilized and the RBS becomes again accessible to the ribosomes for translation (Figure \ref{fig:Figure1}). Therefore, within the cell, RNA thermometers can act as \textit{switches} to adjust the gene-expression according to the environmental temperature\footnote{\textit{dimmers} would probably be more appropriate, if we consider that the transition from the no-translation to translation is typically gradual.}

The known natural RNA thermometers are typically found in pathogenic bacteria where they are used for a host{}-temperature{}-dependent activation of virulence \cite{Hoe1993,Waldminghaus2007,Johansson2002}. RNA thermometers may consist of a single or multiple hairpins, and in general, no accessory factors, such as ligands or proteins, are needed for their functioning \cite{Waldminghaus2007,Johansson2002,Chowdhury2006}. Based on their secondary structure, we can distinguish three distinctive types RNA thermometers: (i) the Repression Of heat Shock protein Expression (ROSE) family (see Figure \ref{fig:Figure1}) \cite{Narberhaus1998}, (ii) the PrfA family \cite{Johansson2002}, and finally, (iii) the FourU family \cite{Waldminghaus2007}.  

Since ROSE RNA thermometers have been better investigated, and more data is available for analysis (32 instances are available of these thermometers), we focused on this class. The ROSE thermometer, consists of 3 RNA hairpins, among which only the hairpin containing the RBS is essential for temperature induced expression \cite{Chowdhury2003}. Therefore, the minimal requirements for designing a functioning ROSE thermometer is that the RNA forms such a hairpin structure (including the RBS), and that the temperature threshold lies within a physiological range. The temperature threshold of an RNA hairpin depends on the thermodynamic stability of the hairpin. In the case of ROSE thermometers, the temperature threshold is in the interval 37C- 42C. It has been shown that the single RBS containing hairpin exhibits a sharper switching behavior from non-translation to translation than the multiple hairpins, thus suggesting that other two hairpins could be present to ``soften'' the switch \cite{Waldminghaus2008}. However, the accompanying hairpins could also have other regulatory functions (e.g. allowing binding of various ligands to adjust the temperature threshold).

To broaden the applicability of RNA thermometers, a tunable threshold within a larger temperature range would highly be desirable, since the known natural RNA thermometers typically work in a very narrow temperature interval, around 37C. 

A strict correlation exists between the Gibbs energy associated to the secondary structure of the mRNA (in particular, the hairpin structure including the RBS), and the translation efficiency of that mRNA \cite{deSmit1990,Nowakowski1997}. An hairpin loop with higher Gibbs energy is likely to have a lower temperature threshold. Therefore, to design a thermometer with a given temperature threshold, it is necessary to somehow modify the Gibbs energy associated to the mRNA secondary structure in order to change its stability. 

Previous studies, trying to elucidate the relation between the translation efficiency and the stability of the temperature sensitive hairpin, are based on the total Gibbs energy of the secondary structure as the main indicator \cite{deSmit1990,Nuepert2008}. In this study, we argue that not just the total Gibbs energy of the secondary structure, but also its distribution along an mRNA strand contributes to determine the sensitivity to temperature: for example, an hairpin loop with a high G-C content in the immediate proximity of the RBS is less likely to work as an RNA thermometer than one with an high A-U or U-G content since G-C base pairs are very stable and would be very difficult to open the hairpin loop. 

Based on this working hypothesis on Gibbs energy distribution, our goal is to design novel RNA thermometers for a given desired temperature. To facilitate this design of RNA hairpins, a software tool has been developed and the temperature controlled gene expression has been described with a model and this model fitted to the experimental data. As a proof of principle, we designed a new RNA thermometer, with a temperature-threshold of 32C, experimentally validated such thermometer and added it as a bio-brick to the registry of standard biological parts. 

\section*{Results and Discussion}

\subsection*{Analysis of Gibbs free-energy distribution at 37C}
\begin{figure}[h]
\centering
\includegraphics[width=0.75\textwidth]{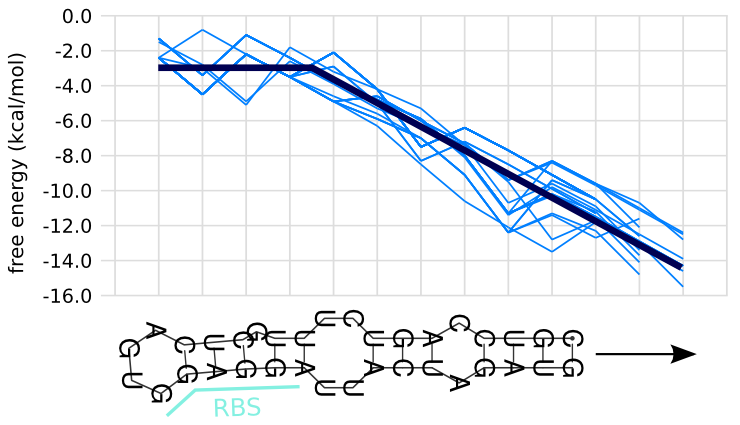}
\caption{Consensus cumulative Gibbs energy distribution found in 32 ROSE{}-type RNA thermometers at 37C, as determined by RNAFold. The light{}-blue lines correspond to the single ROSE RNA thermometers; the dark{}-blue thick line represents  the consensus Gibbs energy distribution.}
\label{fig:Figure2}
\end{figure}

To investigate the relation between the Gibbs free-energy distribution of an RNA strand with a given secondary structure, and its ability to function as an RNA thermometer, a systematic analysis of the Gibbs free-energy distribution of all the 32 known ROSE RNA thermometers with a switching temperature of at 37C has been performed, by means of the software tool RNAFold \cite{Gardner2009,Hofacker2003}. We calculated a cumulative Gibbs free-energy distribution (see the Methods section) for the 32 ROSE RNA, and plotted it along the RNA strand (Figure \ref{fig:Figure2}). From Figure \ref{fig:Figure2}, it can be seen that the cumulative Gibbs free-energy distribution of the different ROSE RNA present a similar characteristic profile (\textit{consensus Gibbs free-energy distribution}): first a plateau of the free-energy, and then a linear reduction. The first part (with a high value of the free-energy) corresponds to the RNA hairpin containing the RBS. This is in agreement with our expectations. In fact, a high value of the free-energy indicates low stability of the hairpin loop (weak base-pairing), and this facilitates the opening of the loop (with the consequent exposure of the RBS for translation) as the temperature increases. 
\subsection*{Design of RNA thermometers with a tunable temperature-threshold}
To design RNA thermometers able to switch at a given desired temperature, we propose a method based on the consensus Gibbs free-energy distribution shown in Figure \ref{fig:Figure2}. 

\begin{figure}
\centering
\includegraphics[width=3.9634in,height=4.2925in]{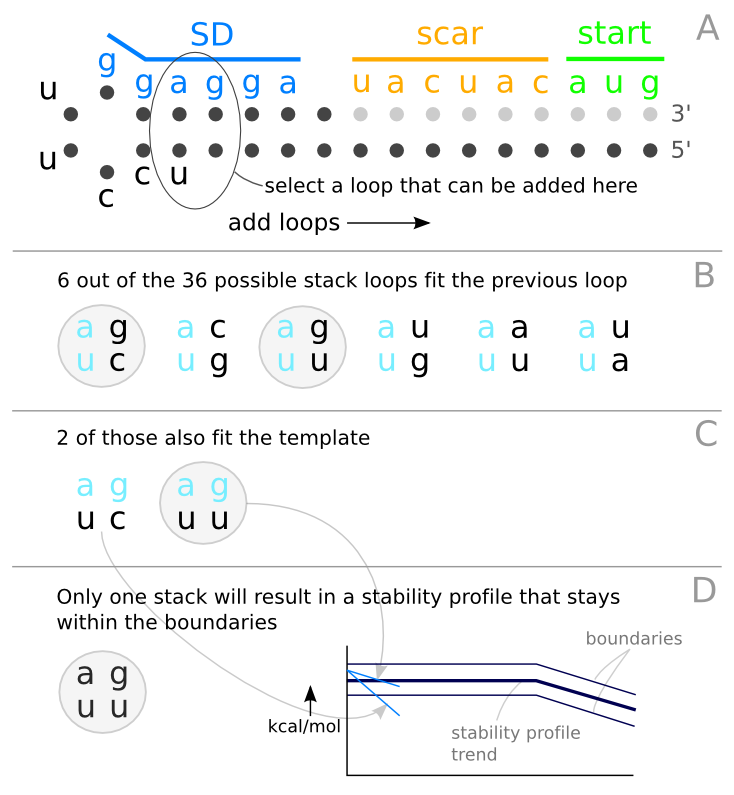}
\caption{Schematic representation of different steps of the proposed hairpin design algorithm. A): A template is selected for the RNA hairpin design. Base pairs are added in loops of 4 nucleotides. The circle indicates where the next loop has to be added. B): All possible loops are generated (36 loops) but only 6 can be added due to the two last bases of the previously added loop (A{}-U in the example). C): The loop to be added has to comply with fixed bases of the given template; only two loops fulfill this requirement. D): Among the feasible loops left (only two in this example), we select the loop that provides the best agreement with consensus distribution of the Gibbs free-energy.}
\label{fig:Figure5}
\end{figure}

The key idea of the proposed design methodology is that the consensus Gibbs energy distribution at a given temperature has to be the same as the one that we have found at 37C for the RNA, to properly work as a thermometer at the new temperature (Figure \ref{fig:Figure2}).  Since Gibbs free-energy distribution is temperature dependent, to keep the same profile, we have to rationally modify the nucleotide sequence. In fact, by altering the nucleotide sequence of the RNA, we can vary the free-energy and thus the stability of the secondary structure. The nucleotides cannot be arbitrarily changed, since there are a number of hard constraints to comply with. Therefore, we formalized the design procedure into a general algorithm to automate the design of RNA thermometers (see method section for more details). Such algorithm has been then implemented as a user-friendly  software tool. The current version of the program can be used to design RNA hairpins of stacked loops. Additonal features such as bulges and internal loops are left as future extensions. The output of the software is a set of RNA sequences that has a desired switching temperature, complying with a desired Gibbs energy distribution at 37C. The user can specify a tolerance band for the consensus profile
(see Figure \ref{fig:Figure5})

\begin{figure}
\centering 
\includegraphics[width=0.55\textwidth]{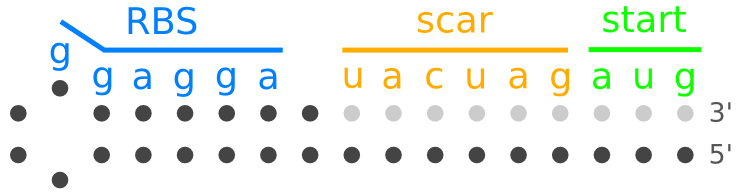}
\caption{The template used as a basis for RNA thermometer design. The scar is a predefined sequence, imposed by the use of standardized parts. The black dots indicate nucleotides that can be substituted; the grey dots indicate nucleotides that are fixed.}
\label{fig:Figure3}
\end{figure}

\begin{figure}
\centering
{\includegraphics[width=4.2972in,height=3.1965in]{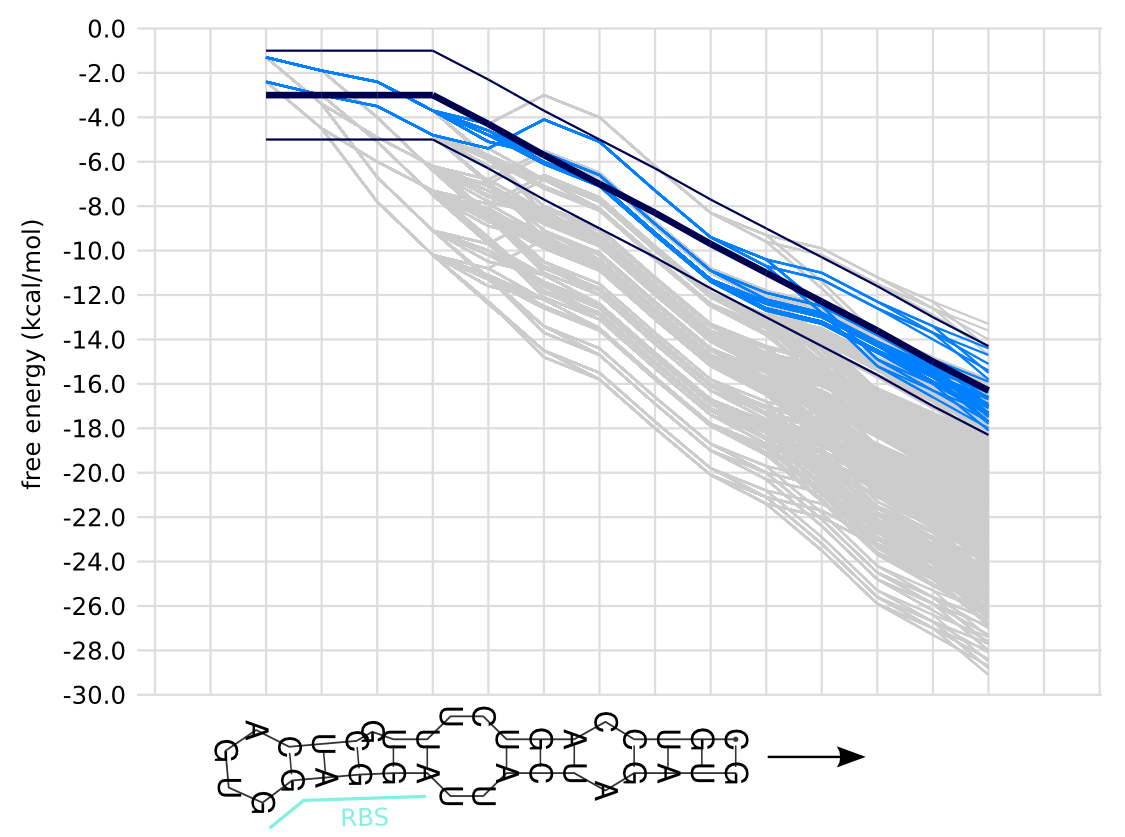}}
\caption{Stability profiles of RNA hairpins designed with the proposed algorithm. The blue lines represent the Gibbs free{}-energy profiles of the hairpins that meet the requirements, namely that are within the tolerance band (delimited by the dark blue thick lines). In light grey are displayed all Gibbs free{}-energy profiles of all possible hairpins loop for the given sequence length. The thick black line is the consensus Gibbs energy distribution. The x{}-axis displays the position of the nucleotide of the hairpin, with on the left side the loop of the hairpin, which is the root of hairpin; whereas on the y-axis the (cumulative) Gibbs free{}-energy has been reported.}
\label{fig:Figure6}
\end{figure}

As a proof of principle, the proposed design method has been applied to design a hairpin with a switching temperature of 32C, starting from a RNA thermometer of the ROSE family (see Figure \ref{fig:Figure3} for the template used in the design). The consensus Gibbs energy distribution of the ROSE family has been used as specification for the design (Figure \ref{fig:Figure2}). 69 feasible solutions were obtained for the RNA sequence, with a tolerance band on Gibbs energy set to 2.0 kcal/mol. The corresponding Gibbs free-energy distributions are depicted in Figure \ref{fig:Figure6}. When the tolerance is set to infinity, the number of feasible hairpin designs explodes to 1536, representing indeed the set of all possible sequences for a hairpin of considered length. For most of these hairpins, the Gibbs energy drops too fast to fit the consensus distribution, resulting in hairpins that are too stable for the RNA thermometer to function at a physiological temperature. This result is due to the algorithm implementing only stacked loops. The implementation of bulges and internal loops in the algorithm is likely to provide more hairpins better fitting the desired consensus distribution. Though the initial design of the thermometer was initially carried out by a trial and error procedure, the developed software tool was then able to reproduce the original design. The novel RNA thermometer that we have designed has been submitted to the registry of standard parts with the identifier BBa\_K115017 (Table \ref{tab:Table1}).

\begin{table}[ht]
\centering
\caption{DNA sequences of the RNA thermometers. ROSE is the naturally occurring RNA thermometer, with some minor alterations to be compatible with the registry of standard parts format. BBa\_K115017 is the thermometer designed to have a temperature threshold of 32C.}

\begin{tabular}{|m{1.0712599in}|m{0.9698598in}|m{4.42406in}|}
\hline
RNA thermometer & Designed threshold T & Sequence\\\hline
ROSE &
37C &
tcccggccgc cctaggggcc cggcggagac gggcgccgga ggtgtccgac gcctgctcgt
acccatcttg ctccttggag gat\\\hline
BBa\_K115017 &
32C &
ccgggcgccc ttcgggggcc cggcggagac gggcgccgga ggtgtccgac gcctgctcgt
ccagtctttg ctcagtggag gat\\\hline
\end{tabular}
\label{tab:Table1}
\end{table}

\subsection*{Experimental analysis of the designed RNA thermometer}
To analyze the designed RNA thermometer (BBa\_K115017), the DNA corresponding to the RNA sequence was synthesized. The RNA thermometer was combined with the \textit{Renilla} luciferase gene in order to conveniently evaluate the performance of the thermometer. Moreover, a second construct with luciferase under control of a regular (i.e. not thermo-sensitive) ribosome binding site was assembled to be used as negative control. 

\begin{figure}
\centering 
\includegraphics[width=0.65\textwidth]{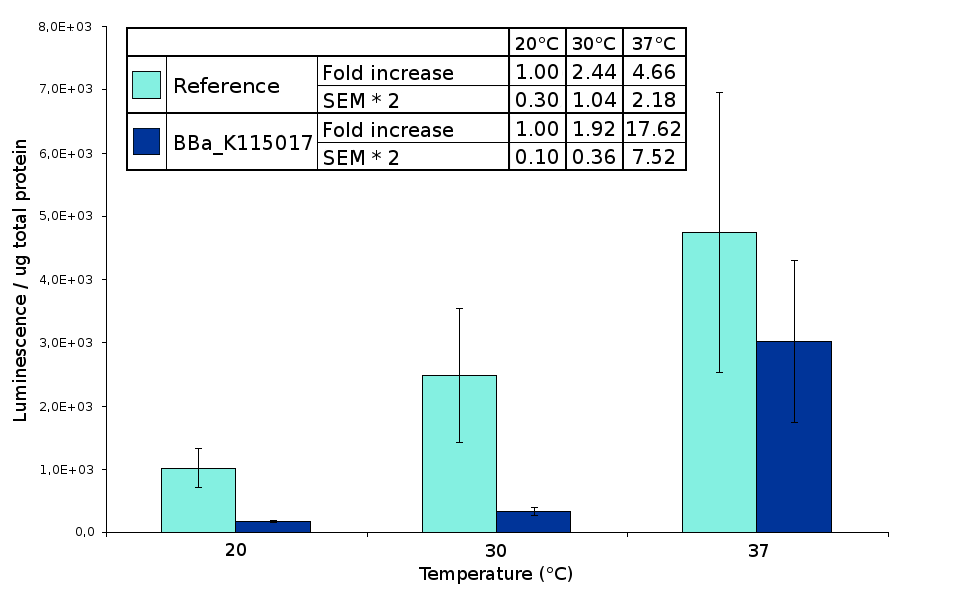}
\caption{Luminescence of the standard part BBa\_K115017 and of the negative{}-control (reference in the figure) without RNA thermometer, measured and corrected for total protein content. Error bars are based on SEM*2 of 3 replicates. Table inset shows the increases normalized for T=20C, and standard deviations}
\label{fig:Figure4}
\end{figure}

Luciferase production of the construct was measured using a luciferase assay on transformed \textit{Escherichia coli} cultures grown at three different temperatures: 20C, 30C and 37C. 1.9{}-fold increase in luminescence in the transition between 20C and 30C was observed (Figure \ref{fig:Figure4}). This is inferior, though of the same order of magnitude, to 2.5{}-fold increase in the control strain. This indicates the RNA thermometer has a repressing effect on translation in comparison to the reference strain at 30C. When temperature was further increased to 37C, the control strain showed again a 1.9{}-fold increment, whereas a 9.2{}-fold increase was observed in the strain with temperature controlled luciferase expression. The observed relative increase is likely to be due to the temperature exceeding the temperature{}-threshold of the RNA thermometer, thus relieving the repressing effect still present at 30C. The findings are consistent with the threshold temperature of 32C for which this RNA hairpin was designed. 

\subsection*{Modeling temperature-dependent enzyme production}
To get more insight on the thermo-regulation of gene-expression, a mathematical model was constructed to describe enzyme production as a function of temperature, based on the measurements of the luciferase production as controlled by the RNA thermometer BBa\_K115017. In this model, we assumed that the mRNA encoding luciferase is constitutively produced, that the production rate is only a function of temperature and we also took into account the inactivation of the enzyme at higher temperatures. The model parameters were fitted to the experimental data via a standard genetic algorithm.

\begin{figure}
\centering
{\includegraphics[trim=4cm 8cm 4cm 8cm, clip,width=0.55\textwidth]{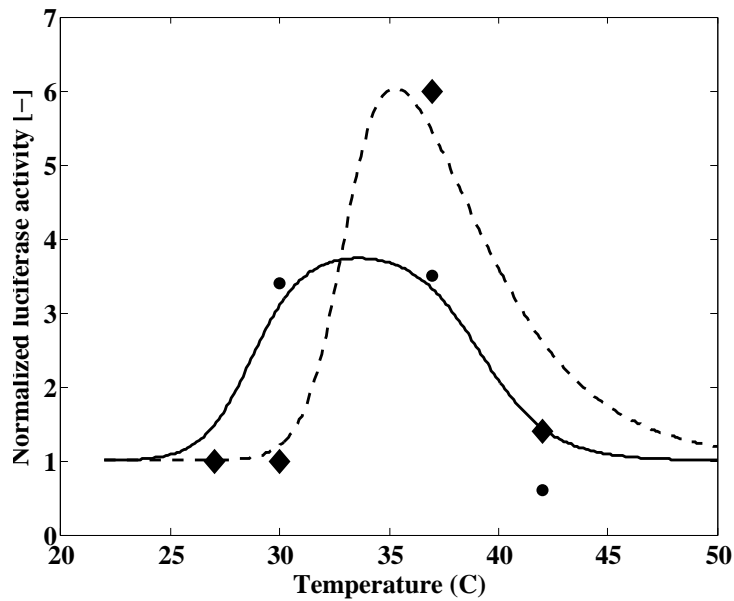}}
\caption{The switching behavior of the model for two different temperatures. The estimated parameters for the model for a temperature threshold at 30C are: $\alpha = 2\times 10^{40}$, m=25, K=28.78, K$_{i}$=39.25 and p=25. The estimated parameters for the model with a temperature threshold at 37C are: $\alpha  = 1.7\times10^{22}$, m=35.72, K=33.11, K$_{i}$=38.2 and
p=13.15}
\label{fig:Figure8}
\end{figure}

We used the model to describe the response of the luciferase activity to the changing temperature for two different constructs: one initially designed to switch on at 30C and the other designed to switch at 37C. The model predictions versus the experimental results for these constructs are plotted in the Figure \ref{fig:Figure8}: we can see that the model predictions are in good agreement with the measured data. Moreover the obtained K parameters (28.78 for the construct designed to switch on at 30C and 33.11 for the construct designed to switch at 37C) indicates that the constructs do switch on near the desired temperatures.

\begin{figure}
\centering 
{\includegraphics[trim=1cm 4cm 1cm 4cm,clip,width=0.65\textwidth]{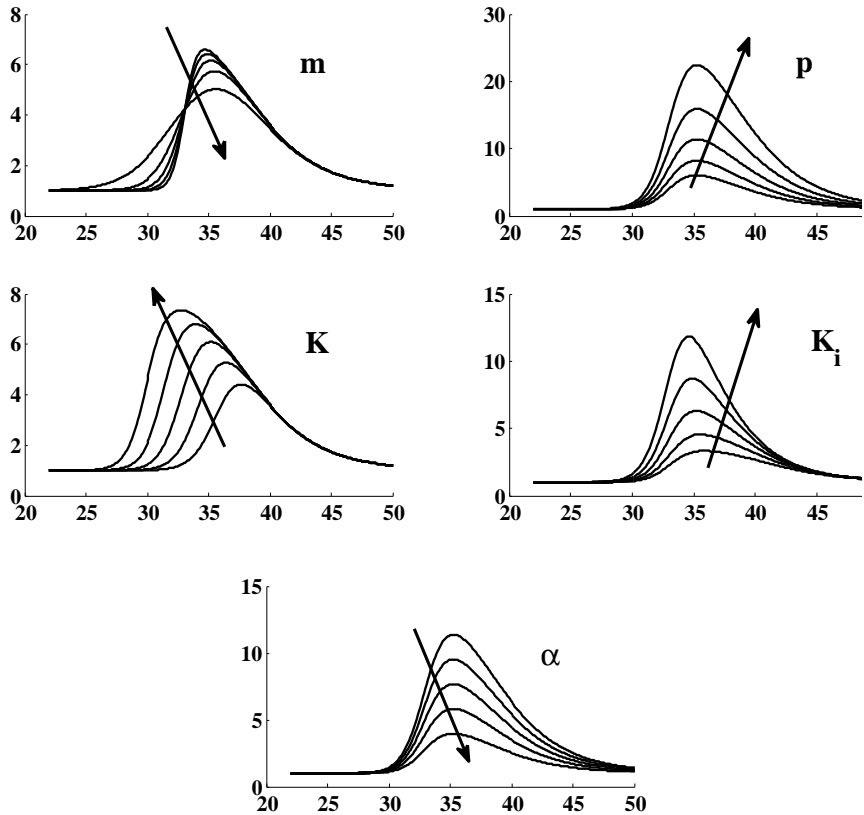}}
\caption{Parameters perturbations reported as the effects of a change in a specific parameters on the normalized enzyme concentration. In each plot, x-axis reports the temperatures and the y-axis the normalized enzyme concentrations, and the arrow points the decreasing direction for the specific plot} 
\label{fig:Figure9}
\end{figure}

\begin{figure}
\centering 
\includegraphics[width=3.5547in,height=3.9992in]{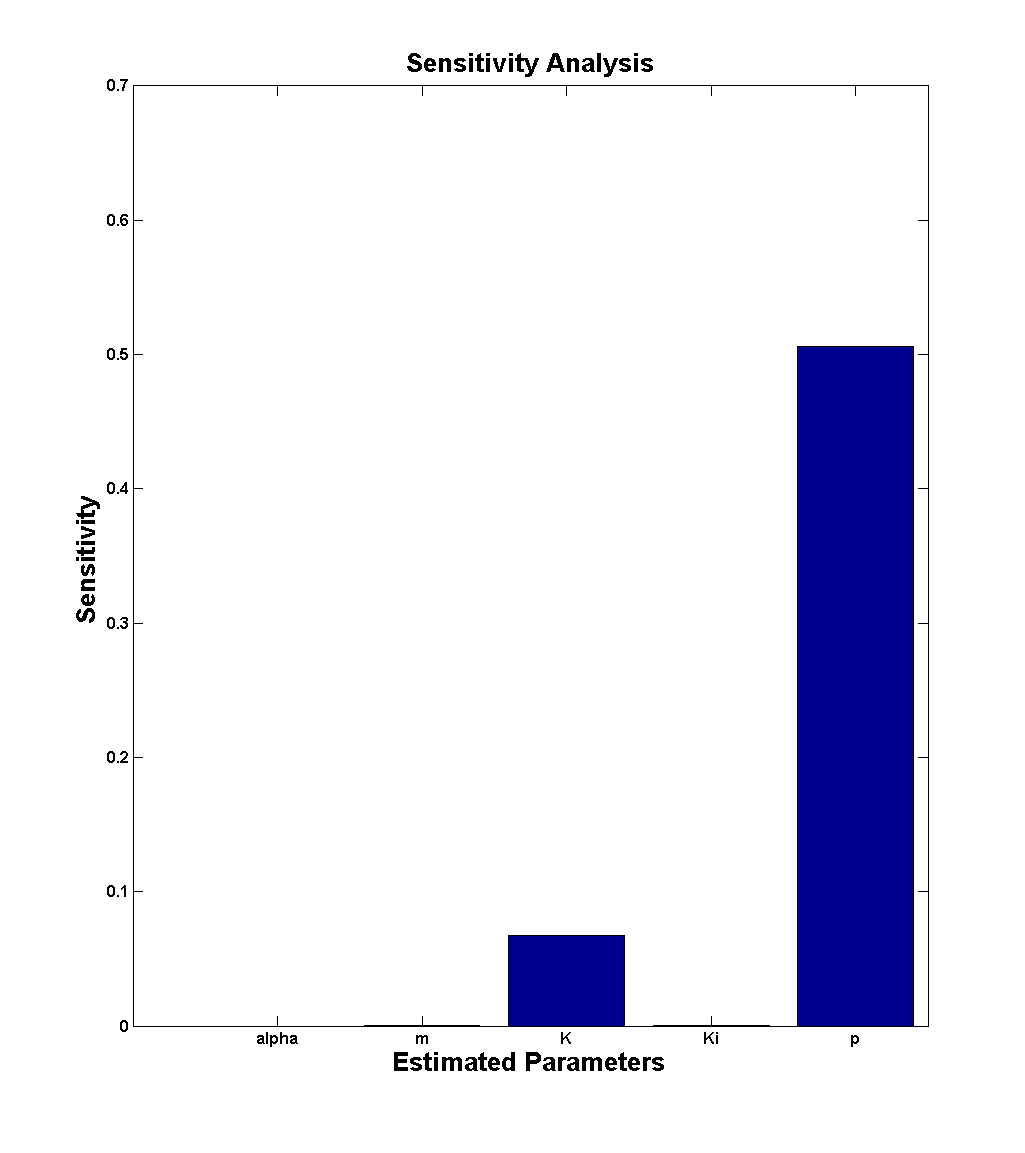}
\caption{Sensitivity analysis of the model. The model is sensitive to K and P.}
\label{fig:Figure10}
\end{figure}

We explored the constructed model, and presented the effects of changes in different parameters on the normalized enzyme concentration in Figure~\ref{fig:Figure9} to get more insights on the dynamics the model. From these simulations, $m$ describes the hardness of the switch (bigger the value of $m$, harder the switch), $K$ determines the switch-on temperature. On the inhibition side, the hardness and the temperature of the switch-off is described by a non-obvious combination of $p$ and $K_i$. Lastly, sensitivity analysis results depicted in Figure \ref{fig:Figure10} showed that the parameter $p$, explaining the hardness of the switch-off at elevated temperatures,  is the parameter in the model with the highest sensitivity. This is understandable as it is know that at elevated temperatures, the loss of activity of the enzyme is quite abrupt. Also, the model is quite sensitive to the level of $K$, but it is almost five times less sensitive than p. Moreover, from the results, the sensitivity of the model to the level of m and K$_{i}$ is very low. Finally, the parameter $\alpha$ which represents the maximum production rate relative to the basal production rate has expectedly very low sensitivity.

\section*{Conclusion}
In this work, we have proposed a novel and general methodology for the design of RNA thermometers with a tunable temperature threshold. The methodology is based on the idea that an RNA strand can work as a thermometer if its secondary structure has a well defined Gibbs free-energy profile. Such consensus distribution profile has to guarantee that the RNA forms a hairpin occluding the ribosome binding site (as long as the temperature is below a certain threshold); and that the hairpin unwinds as soon as the temperature exceeds that threshold. Our algorithm automatically determines how to select the nucleotides of the hairpin to achieve the desired consensus distribution for a certain temperature.

The proposed method has been implemented in a user{}-friendly, freely available, software tool, and this tool has been applied to the design of an RNA thermometer BBa\_K115017 with 32C as temperature threshold. This sequence of BBa\_K115017 was derived from a RNA of the ROSE family with a temperature threshold between 37C and 42C. The ]{The proposed method has been implemented in a user-friendly, freely available, software tool, and this tool has been applied to the design of an RNA thermometer BBa\_K115017 with 32C as temperature threshold. This sequence of BBa\_K115017 was derived from a RNA of the ROSE family with a temperature threshold between 37C and 42C. The experimental results show a temperature threshold between 30C and 37C that is in agreement with the expected value of 32C. Furthermore, a model has been built to describe the behavior of the RNA thermometer. The model with the estimated parameters is able to reproduce the observed experimental results. 

In this work, we have shown that RNA thermometers can be used as an alternative way to post-trans\-crip\-tio\-nal\-ly regulate gene expression. In perspective, this could reduce the costs in large{}-scale fermentation and in downstream processing since no induction chemical needs to be added. However to practically achieve this goal, further research is needed; 1) the automated design method can be enhanced by allowing internal loops and bulge loops in the hairpin, thus increasing the space of possible solutions; 2) the proposed methodology should be validated on several different RNA thermometers; 3) the measurements of the luciferase levels should be performed with an higher resolution in order to better characterize the switching of the RNA thermometers; 4) different recombination techniques (other then the one used for BioBricks), should be investigated in order to have different patterns of fixed nucleotides in the initial RNA template (for example short tag protein could be used).

\section*{Methods}
\subsection*{Gibbs free-energy distribution analysis}
The Vienna RNA Package was used to analyze the Gibbs energy distribution \cite{Hofacker2003}. The 32 ROSE RNA thermometers used for the analysis were retrieved from the Rfam database \cite{Gardner2009}. RNAfold was used to predict the secondary structures of the 32 sequences and RNAeval was
used to evaluate the Gibbs energies of these structures.  

\subsection*{Design of tunable RNA-thermometers}
The key assumption for the design of RNA thermometers at different temperatures is that the Gibbs free-energy distribution associated to the RNA secondary structure has to have a specific profile, what we called the \textit{consensus distribution}. Such a distribution has been determined from the analysis of the 32 RNA thermometers from the ROSE family at 37C. Since the Gibbs free-energy distribution is temperature dependent, in order to preserve the same profile, the nucleotide sequence have to be properly adjusted. However, the nucleotides cannot be arbitrarily changed, since there are some hard constraints to comply with. First, the mutations should not allow the RNA to fold into a nonfunctional secondary structure i.e. the RBS should be occlude to prevent the translation. Second, there are some functional constraints to the possible mutations: e.g. the start codon and the ribosome-binding site cannot be altered. Finally, when using parts in BioBrick format, some bases of the sequence are also fixed: they are the outcome of the standard recombination techniques. In fact, if we use these recombination techniques to assemble multiple standard parts into a composite part, we end up with a construct with a so called \textit{scar} consisting of 6 fixed bases between each two parts assembled. Since the RBS of the thermometer and the protein coding sequence are next to each other, the scar ends up as being part of the functional sequence of the RNA thermometers, and its 6 bases cannot be manipulated. The template RNA hairpin in Figure \ref{fig:Figure3} summarizes these three constraints (starting codon, Shine-Dalgarno region, and scar). The black dots represent the nucleotides that can be manipulated, whereas the grey dots stand for the fixed nucleotides. 

We devised an algorithm for the design of RNA thermometers that takes into consideration all such constraints. For computational efficiency, the algorithm modifies the RNA sequence by iteratively adding a loop at a time, from left to right starting from the initial hairpin loop (Figure \ref{fig:Figure3}). The main steps of the algorithm are: 

\begin{enumerate}
\item Base pairs are modified in groups of 4 starting from left to right, on a template RNA hairpin. Base pairs have to be added in loops of 4 since this is the minimum amount of base pairs of which the Gibbs energy can be calculated.
\item At each step (i.e. addition of a new loop), all possible loops corresponding to the selected 4 nucleotides are generated, but only those consistent with the two last bases of the previously added loop are retained.
\item Only the loops fulfilling the constraints (fixed nucleotides) are selected. 
\item The Gibbs free-energy distribution associated to the added loop is determined and only the loop that provides the best agreement with consensus distribution of the Gibbs free{}-energy is
selected.

\end{enumerate}

A schematic representation of the developed algorithm is shown in Figure \ref{fig:Figure5} for a particular template RNA. For the considered RNA thermometer, there are 36 possible loops, consisting of 4 base pairs that can be added at each step. For the design step shown in Figure 5, only 6 loops can be added for the compatibility with the last two bases of the previously added loop (A-U). Only two out of these 6 loops are consistent with the fixed bases of the given template. And finally only one loop has a Gibbs free-energy compatible with the desired consensus distribution. 

\subsection*{Software application}
The RNA hairpin designer software tool has been written in Java 6 using Eclipse as software development tool. The Gibbs energies used to determine the Gibbs energy distribution within the hairpin structure are the same as in RNAeval. The tool has been developed to obtain hairpin structures with a Gibbs free-energy distribution that approximates a given consensus distribution, (with a certain tolerance that can be also specified by the user) at a certain temperature. This can be achieved, in general, by adding bulge loops, internal loops, and stack loops to the hairpin structure (Figure \ref{fig:Figure7}). In the current version of the software, only stacked loops have been implemented. The loop of the hairpin is considered as the root of a search tree; all base pair loop additions that meet the requirements are then child nodes. In this way, all possible hairpins will be designed recursively, and can be found at the end of the branches of the search tree. 
\begin{figure}
{\centering 
\includegraphics[width=4.0071in,height=2.2201in]{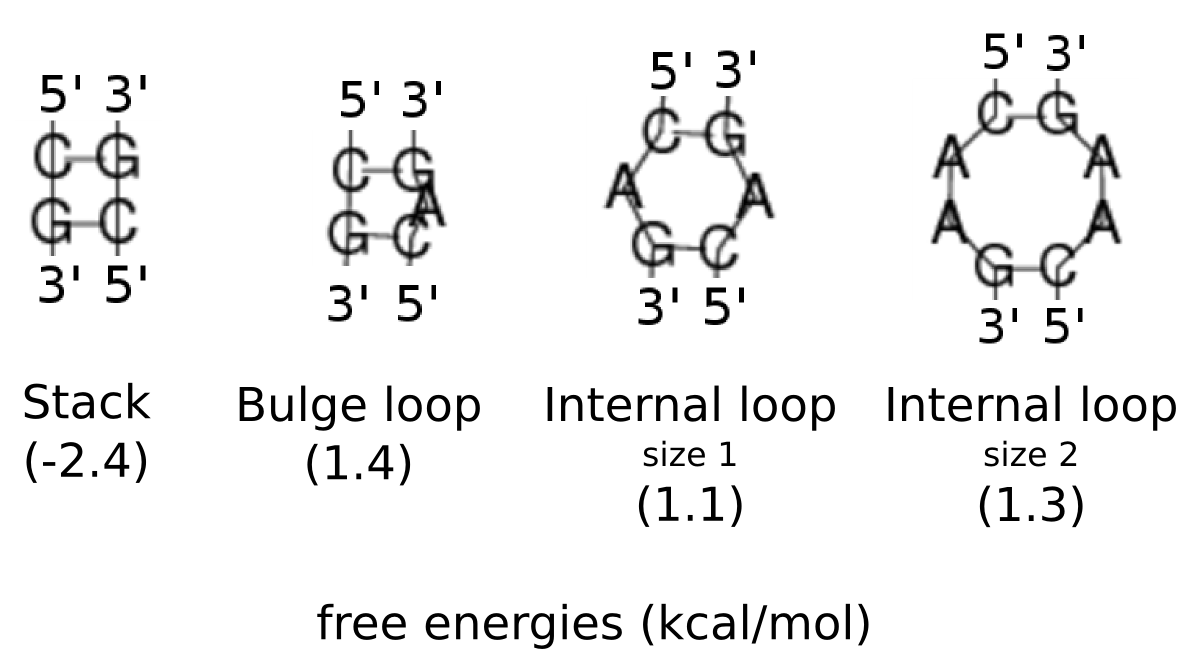}
\par}
\caption{The possible structures of loops that can be added to a hairpin. A hairpin structure consists of a hairpin loop and a stem. The stem is build out of different types of loops each having their own Gibbs free{}-energy. The sum of all these Gibbs energies provides the total Gibbs free{}-energy of the hairpin structure. The values of the Gibbs free{}-energy given in the figure refer to the specific base pairs depicted: that is, the same structures but with different base pairings would have in general different values.}
\label{fig:Figure7}
\end{figure}

The software allows plotting in a easy way the cumulative Gibbs energy distributions within the temperature sensitive hairpin structures. The software can be freely downloaded \cite{igemTUDelft}. 

\subsection*{Constructs}
The DNA construct used as negative control consists of a pBAD derived promoter (BBa\_R0080), an RBS based on the Elowitz repressilator (BBa\_B00340) \cite{Elowitz2000}, luciferase, based on luciferin 2-monooxygenase from \textit{Renilla reniformis }(BBa\_J52008) followed by two transcription terminators, T1 from \textit{Escherichia coli} rrnB and TE from bacteriophage T7 (BBa\_B0015). The construct for the RNA thermometer has the same components as the negative control, except for the Elowitz RBS that has been replaced by the designed temperature-sensitive RNA BBa\_K115017 (see Table \ref{tab:Table1}). The DNA corresponding to the RNA thermometer was synthesized by \textit{GeneArt}. Assembly of all constructs was performed by using 3A assembly as suggested in \cite{Openwetware}. The control construct was in a pUC19-derived pMB1 vector containing ampicillin and kanamycin resistance (BBa\_pSB1AK3). The thermometer constructs were in a pUC19-derived pMB1 vector containing ampicillin and tetracycline resistance (BBa\_pSB1AT3). 

\subsection*{Luciferase measurements}
For measuring the temperature induced luciferase production, vectors were transformed to TOP10 cells (\textit{Invitrogen}). The strain with the RNA thermometer BBa\_K115017 and control strain were all grown at 20C, 30C and 37C in Lysogeny Broth (LB) medium under ampicillin selection (0.1 mg/ml) \cite{Bertani1951}. When OD's were around 0.9, the cultures were pelleted by centrifugation for 5 minutes at 10000 rpm in a table top centrifuge, re-suspended in PBS, and lysed by sonication on ice for 2 times 15 seconds with a 15 second cool down. Samples were centrifuged for 2 minutes at 14000 rpm, and supernatant samples were stored at -20C. Luciferase measurements were performed using the \textit{Renilla Luciferase Assay System} (\textit{Promega}). Measurements were performed on a BioTek Gen5 plate reader. Luminescence was measured from 2 to 12 seconds after assay buffer addition. Total protein concentration was determined using a \textit{BCA-assay kit} (\textit{Uptima}). All luciferase measurements were corrected for total protein content of the sample. 

\subsection*{Modeling temperature-induced luciferase production }
The evolution over time of an enzyme concentration can be described by equation \ref{eq:MassBal} 

%
\begin{align}
	\dfrac{d\left[e\right]}{dt}= V_p - V_c
	\label{eq:MassBal}
\end{align}

where $\left[e\right]$ is the enzyme concentration, $V_p$ is the production rate and $V_c$ is the consumption or degradation rate for that enzyme. Since we design a thermo-switch, we described the enzyme production rate by a Hill-type kinetics, and as for the degradation of the enzyme, linear kinetics is chosen, such that:

\begin{align}
		V_p = V_{\max} \dfrac{T^m}{T^m+K^m} + V_0 \qquad	\qquad V_c = K_d \left[e\right]
\end{align}		
	
%
%

where T is the temperature, K is the affinity coefficient, m is the Hill coefficient, V$_{\max}$ is the maximum rate of production, $_{0}$ is a basal production rate, and finally K$_d$ is the degradation constant. At steady state, the production rate equals the consumption rate in equation \ref{eq:MassBal}, therefore the enzyme concentration can be calculated as a function of temperature

\begin{align}
	[e] = \dfrac{V_{\max}}{K_d}\dfrac{T^m}{T^m+K^m} + \dfrac{V_0}{K_d}
\end{align}
%

In preliminary experiments (unpublished), we found out that a culture growing at a temperature above 40C presents a reduced luminescence. This is likely due enzyme inactivation induced by the heat. In order to model the enzyme inactivation, we included an additional term into the model. 
%
%
\begin{align}
	\dfrac{\left[e\right]}{\left[e_0\right]} = \alpha \dfrac{T^m}{T^m+K^m}\dfrac{1}{T^p+K^p} + 1
	\label{eq:efinal}
\end{align}
where 
$\alpha = \dfrac{V_{\max}}{V_0}$ and $e_0 = \dfrac{V_0}{K_d}$ .

For the estimation of the model parameters, the sum of the squared error between the model prediction and experimental data was minimized by using a genetic algorithm (GA) from Matlab GA toolbox.. The main GA parameters were so set: crossover rate = 0.8, population size = 100 , termination tolerance = 1e$^{-10}$ and maximum number of generations = 5000.

In order to investigate the effect of the different model parameters on the enzyme activity, a sensitivity analysis was performed by analytically calculating the Jacobian of the normalized enzyme activity equation (Eq. \ref{eq:efinal}) and then evaluating the Jacobian for the estimated parameter set.

\section*{Competing interests}
The authors declare they have no competing interests
\section*{Authors' contributions}
BAB has devised the RNA thermometer design method and has written the corresponding software application.. RH and FE modeled RNA thermometer controlled expression. OMJAS, SMF and RJJJ have carried out the experimental work.  JAK , MJLG, IEN and DB supervised the project. OMJAS wrote the initial draft of the manuscript and IEN and DB revised it. All authors read and approved the final manuscript. All authors read and approved the final manuscript.

\section*{Acknowledgements}
This research has been initiated as part of the international genetically engineered machine competition organized by the Massachusetts Institute of Technology; therefore, we wish to thank the iGEM founders, organizers, and community for providing a supportive environment for conducting synthetic biology research. This project has been financed by a grant of the Department of Biotechnology, Delft University of Technology: in particular, we would like to thank Han de Winde for his unconditional support. We also acknowledge the financial support from the \textit{Kluyver Center for Genomics of Industrial Fermentation, NGI, NWO}, from the Department of Biotechnology Bionanoscience of the Delft University of Technology, and from Royal DSM. We acknowledge Marijke Luttik, Loesje Bevers, Laura Koekkoek and Esengl Yildirim for supporting us in the lab work. We also acknowledge Susanne Hage for the biological safety permit. Finally, we thank Ali Mesbah for his advises on the modeling part.

%


\end{document}